\documentclass[conference]{IEEEtran}
\IEEEoverridecommandlockouts
\usepackage{cite}
\usepackage{amsmath,amssymb,amsfonts}
\usepackage{algorithmic}
\usepackage{graphicx}
\usepackage{textcomp}
\usepackage{xcolor}

\def\BibTeX{{\rm B\kern-.05em{\sc i\kern-.025em b}\kern-.08em
    T\kern-.1667em\lower.7ex\hbox{E}\kern-.125emX}}

\begin{document}

\title{Hybrid Quantum Machine Learning Assisted Classification of COVID-19 from Computed Tomography Scans
\thanks{© 2023 IEEE.  Personal use of this material is permitted.  Permission from IEEE must be obtained for all other uses, in any current or future media, including reprinting/republishing this material for advertising or promotional purposes, creating new collective works, for resale or redistribution to servers or lists, or reuse of any copyrighted component of this work in other works.

This work was partially funded by the BMWK project \textit{PlanQK (01MK20005F / 01MK20005I / 01MK20005P)}}
}

\author{\IEEEauthorblockN{Leo Sünkel}
\IEEEauthorblockA{\textit{LMU Munich} \\
leo.suenkel@ifi.lmu.de}
\and
\IEEEauthorblockN{Darya Martyniuk}
\IEEEauthorblockA{\textit{Fraunhofer FOKUS}\\
darya.martyniuk@fokus.fraunhofer.de}
\and
\IEEEauthorblockN{Julia J. Reichwald}
\IEEEauthorblockA{\textit{Smart Reporting GmbH}\\
j.reichwald@smart-reporting.com}
\and
\IEEEauthorblockN{Andrei Morariu}
\IEEEauthorblockA{\textit{Smart Reporting GmbH}\\
a.morariu@smart-reporting.com}
\and
\IEEEauthorblockN{Raja Havish Seggoju}
\IEEEauthorblockA{\textit{Fraunhofer FOKUS}\\
raja.havish.seggoju@fokus.fraunhofer.de}
\and
\IEEEauthorblockN{Philipp Altmann}
\IEEEauthorblockA{\textit{LMU Munich}\\
philipp.altmann@ifi.lmu.de}
\and
\IEEEauthorblockN{Christoph Roch}
\IEEEauthorblockA{\textit{LMU Munich}\\
christoph.roch@ifi.lmu.de}
\and
\IEEEauthorblockN{Adrian Paschke}
\IEEEauthorblockA{
\textit{Freie Universität Berlin}\\
\textit{Fraunhofer FOKUS}\\
adrian.paschke@fokus.fraunhofer.de}
}

\maketitle 

\begin{abstract}
\textbf{Practical quantum computing (QC) is still in its infancy and problems considered are usually fairly small, especially in quantum machine learning when compared to its classical counterpart. Image processing applications in particular require models that are able to handle a large amount of features, and while classical approaches can easily tackle this, it is a major challenge and a cause for harsh restrictions in contemporary QC. In this paper, we apply a hybrid quantum machine learning approach to a practically relevant problem with real world-data. That is, we apply hybrid quantum transfer learning to an image processing task in the field of medical image processing. More specifically, we classify large CT-scans of the lung into COVID-19, CAP, or Normal. We discuss quantum image embedding as well as hybrid quantum machine learning and evaluate several approaches to quantum transfer learning with various quantum circuits and embedding techniques.}
\end{abstract}


\section{Introduction}
Quantum computing (QC) is an ever-growing field with applications spanning a wide range of domains including finance \cite{herman2022survey,egger2020quantum}, chemistry \cite{kassal2011simulating,cao2019quantum}, simulation \cite{brown2010using}, machine learning \cite{schuld2015introduction,biamonte2017quantum} and optimization \cite{farhi2014quantum}. While attention for this field is rising rapidly, practical QC is itself in its infancy and the capabilities of current and near-term quantum computers are limited. Thus, the current phase of QC is famously referred to as the noisy intermediate scale quantum era (NISQ-era) \cite{preskill2018quantum}. 
Yet, QC has the potential to solve some computational problems faster than classical computers, for example providing an exponential \cite{shor1999polynomial} or quadratical \cite{grover1996fast} speedup. Quantum machine learning (QML) currently is a popular domain amongst researchers in which many hope to find a quantum advantage or speedup. 
In this paper, we apply a QML assisted approach to a real-world problem from the medical domain, the classification of COVID-19 from CT scans of the lung. However, as these CT scans are images with a large number of features and the capacity of current QC-hardware is fairly limited, we employ a hybrid approach, i.e., an approach that consists of a classical as well as a quantum part, instead of a purely quantum approach.  
The paper is structured as follows. In Section \ref{sec:related_work} we briefly discuss related work. 
We introduce the necessary background in Section \ref{sec:background}, i.e., we discuss the topic of classification of medical images in the context of the detection of COVID-19 and give an overview of various quantum embedding methods including image specific encodings such as FRQI as well as more general encodings like angle encoding. We conclude the background section with an overview of quantum transfer learning (QTL). In Section \ref{sec:methods} we discuss QTL approaches and quantum circuits used in our experiments while the experimental setup and results are presented in Section \ref{sec:experimental_setup_and_results}. We discuss the results in Section \ref{sec:discussion} and give a conclusion in Section \ref{sec:conclusion}.

\section{Related Work}\label{sec:related_work}
 Hybrid algorithms are a popular substitute for algorithms that are designed to solely run on quantum computers. One popular hybrid algorithm is the classical-quantum transfer learning approach proposed by Mari et al. in \cite{mari2019transfer}. In this approach, a pre-trained classical neural network is combined with a variational quantum circuit (VQC) that can then be used for classification tasks. The authors furthermore propose their dressed quantum circuit (DQC), a VQC that is comprised by two classical layers, i.e., pre- and post-processing layers. This approach is particularly suitable for large image classification. An alternative to this QTL approach was proposed in \cite{icaart23}. Pramanik et al. propose a hybrid quantum-classical pipeline for the problem of crack detection in images, this task includes image classification and segmentation \cite{pramanik2022quantum}. Otgonbaatar et al. \cite{otgonbaatar2022quantum} employ quantum-transfer learning to image classification, an approach similar to Mari et al.. Azevedo et al. apply a hybrid quantum transfer learning approach to breast cancer detection, i.e., a binary classification problem where the task is to classify images (mammograms) into either malignant or benign \cite{azevedo2022quantum}. The proposed method also uses the dressed quantum circuit model proposed by Mari et al. In \cite{majumdar2023histopathological} a hybrid quantum transfer learning approach is applied to histopathological cancer detection. The authors evaluate several classical networks and VQCs. Landman et al. propose and evaluate two quantum based methods for image classification in the medical domain \cite{landman2022quantum}. Umer et al. employ a QTL approach based on Mari et al. for COVID-19 classification using X-ray images \cite{umer2022integrated}. In this paper, we apply two different variants of QTL. The first approach uses the DQC mentioned above. In this approach, the quantum part of the algorithm is trained alongside the classical pre- and post-processing layers, though this makes it difficult to determine the exact impact of the quantum circuit. We therefore also evaluate a second approach in which we remove both classical layers usually used in other QTL approaches and only optimize the parameters of the quantum circuit, thereby strictly separating the classical and quantum parts of this hybrid approach. We then evaluate and compare both approaches.

\section{Background}\label{sec:background}
Before we discuss our approach to the problem of classifying large CT-scans with the assistance of QC, we will introduce the necessary background information in this section. We start by discussing the classification of images from a medical perspective with regards to COVID-19. We then briefly introduce quantum embedding techniques, general approaches as well as image-specific. We continue by briefly discussing QML and QTL.

\subsection{Medical image classification}
The COVID-19 pandemic posed significant challenges to healthcare systems around the world. Early and accurate diagnosis of COVID-19 infection was critical, both for individual patients who benefit from early treatment and for the general population, which could be protected by early isolation of infected patients. 

While rRT-PCR was widely considered the gold standard for detecting COVID-19 infection \cite{Tahamtan2020}, imaging techniques such as X-ray and CT scans of the chest have also proven useful. Although X-rays are widely available and more economical, the sensitivity is lower than CT scans, thus chest CT is the most accurate method of assessing the tendency and severity of COVID-19 infection \cite{Yang2020}. Moreover, CT scans are more sensitive in the early stages of COVID-19 infection than rRT-PCR, which has been found to have a high rate of false-negative results \cite{Fang2020}. Patients with the suspicion of COVID-19 pneumonia will undergo a CT scan of the chest, without contrast enhancement, however since COVID-19 infection is linked with an increased risk of acute thromboembolic events, if there is a suspicion of pulmonary embolism, a CT angiography will be performed.

COVID-19 infection can range from normal chest CT findings to typical COVID-19 pneumonia or, less frequently, cavitating lesions. The typical COVID-19 pneumonia findings in early stages in CT scans include Ground-Glass Opacities~(GGO) that are characterized as hazy areas with slightly increased density in lungs without obscuration of bronchial and vascular margins, which may be caused by the partial displacement of air due to partial filling of airspaces or interstitial thickening \cite{Ye2020}. The distribution of the GGOs is usually peripheral, multifocal, and bilateral. A common finding within GGO areas is the widening of the vessels. Crazy paving patterns can also be visible during early stages, which demonstrate thickened interlobular septa (interlobular thickening) and intralobular lines with superimposition on a GGO background, resembling irregular paving stones. CT images of COVID-19 patients in later stages can also show signs of consolidation, i.e. alveolar air being replaced by pathological fluids, cells, or tissues, manifested by an increase in pulmonary parenchymal density that obscures the margins of underlying vessels and airway walls. The reverse halo signs, which represent peripheral consolidation with GGO in the center can be also visible in COVID-19 patients. In later stages, the consolidations and GGOs reduce and gradually disappear, while the reticular pattern becomes more evident.
Yet all these image findings can occur in other lung diseases equally and pathognomonic CT findings of COVID-19 have not been identified \cite{Sahu2020}, which makes it difficult to distinguish COVID-19 from other diseases leading to similar findings in lung CT scans.

There are multiple classifications and grading systems for COVID-19 lung findings. CO-RADS is based on the CT findings and provides a level of suspicion of COVID-19 pneumonia (CO-RADS 1 represents a normal chest CT and CO-RADS 5 represents typical COVID-19 findings and CO-RADS 6 a positive rT-PCR test). Additionally, the CT severity score (CTSS) assesses the affected region within each lobe, as a percentage and assigns a grade according to that, which correlated with clinical parameters help clinicians choose the most appropriate treatment course. Both systems have proven useful in clinical practice, the first in diagnosing COVID-19 pneumonia and the second in establishing the severity.

Researchers around the world have therefore developed machine learning algorithms based on CT scans to properly detect and predict COVID-19 infection \cite{Roberts2021}. But these existing models have methodological flaws and underlying biases and the conventional AI-assisted analysis of CT data sets of lungs of COVID-19 patients currently takes about ten minutes \cite{PMSR22}. This makes the current pipeline way too slow for real-time diagnosis in everyday clinical practice. Therefore, further research is needed on how to improve the quality and ultimately the clinical utility of these models with the goal to have an automatic classification of lung CT images into "healthy," "COVID-19 pneumonia," or "other findings".

\subsection{Quantum (image) embedding}
Current quantum computers contain relatively few and error-prone qubits, and error-correction is not feasible. This has severe consequences for algorithm development in many ways. For instance, only algorithms with a low depth, i.e., a low number of gates, and few qubits can be executed on actual quantum hardware. This limits the number of classical features that can be efficiently embedded through quantum state preparation, an important procedure in particular for QML applications. Images contain a large number of features, especially CT-scans may have many thousands, and embedding all features efficiently presents a difficult, if not impossible challenge. In QC there exist many embedding strategies or procedures, general as well as image-specific, and we will introduce a selection next.

\subsubsection{FRQI}
 The Flexible Representation of Quantum Images (FRQI)~\cite{le2011flexible} enables representation of a $2^n \times 2^n$ classical image as a quantum state $|I\rangle$ by encoding colors and its corresponding position in the image using control rotation matrices and $2n + 1$ qubits:
 \begin{equation}
|I\rangle = \frac{1}{2^n}\sum_{i=0}^{2^{2n}-1} (\cos \theta_i |0\rangle + \sin \theta_i |1\rangle) \otimes |i\rangle, \label{eq_1}
 \end{equation}
 where 
 $|i\rangle$, $i = 0, 1,..., 2^{2n}-1$ is a $2^{2n}$-D computational basis quantum state that encodes position of the corresponding pixel, and  $\theta = (\theta_0, \theta_1,...,\theta_{2^{2n}-1})$ encodes the color values of each  pixel. For example, in a 2$\times$2 classical grayscale image, $i$~would take values  $00$, $01$, $10$, and $11$, and  $\theta_i$ be within the range~$[0, \frac{\pi}{2}]$ with $0$ for black and $\pi/2$ for white. Since FRQI uses only one qubit for color information, it is limited to represent pixel-wise complex operations.
 
\subsubsection{NEQR}
The Novel Enhanced Quantum Representation (NEQR)~\cite{zhang_neqr_2013} approach was introduced as an improvement of FRQI. To overcome the limitation of FRQI, it uses $n$ qubits to store the color values to encode classical grayscale images in a normalized superposition.
According to NEQR, the classical $2^n \times 2^n$ image can be represented as a quantum state $|I\rangle$ with $2n + n$ qubits as follows: 
 \begin{equation}
 |I\rangle = \frac{1}{2^n}\sum_{i=0}^{2^{2n} -1}\otimes_{i=0}^{n-1}|C_{i}^j\rangle |i\rangle,  \label{eq_2}
\end{equation}
 where the binary sequence $C_{i}^j$, $j = 0, 1, ..., n-2, n-1$ encodes the color information $f(i)$ of the corresponding pixel $i$.

Enhanced versions of the NEQR such as Improved Novel Enhanced Quantum Representation~(INEQR)~\cite{Jiang_Wang_2014} and generalized model of NEQR~(GNEQR)~\cite{GNEQR} have been introduced in the literature. Moreover, there have been  research results on further encoding schemes in the filed of quantum image processing, e.g., the Quantum Indexed Image Representation~(QIIP)~\cite{QIIP}, the Quantum Block Image Representation~(QBIR)~\cite{QBIR} and  the Double Quantum Color Images Representation Model~(DQRCI)~\cite{Wang2019DoubleQC} just to name a few. However, in terms of hardware resources required for its realization, these methods are unfeasible on the NISQ-devices when it comes to encode chest CT-images.
\subsubsection{Amplitude embedding}
Amplitude embedding encodes a normalized feature vector $x = (x_1, x_2, ..., x_{2^n})$, $\sum{|x_i| = 1}$ of a classical $n \times n$ image into the amplitude vector of $log_{2}n$ qubits:
 \begin{equation}
 |I\rangle = \otimes_{i=0}^{2^n}x_i|i\rangle,  \label{eq_3}
  \end{equation}
where $x_i$ is the $i$-th feature and $|i\rangle$ is the $i$-th vector of the computational basis~\cite{SchuldQMLBook}. Note that amplitude embedding can be used to encode arbitrary features, i.e., it is not restricted to images.

\subsubsection{Angle embedding}
Angle embedding is widely used in QML applications. In this method, each feature $x_i$ is encoded as a rotation parameter in the $RY(x_i)$ or $RX(x_i)$ rotation gate, which is applied on a qubit in some initial state, e.g., $|0\rangle$:
 \begin{equation}
 |I\rangle = \otimes_{i=0}^{N}R_k(x_i)|0\rangle , \label{eq_4}
  \end{equation}
where $k=\{X,Y\}$ is the rotation axis and $N = n*m$ is the number of features of a classical  $n \times m$ image.
Thus, angle embedding requires $N$ qubits and a single gate per qubit for its realization.  

An advanced version of this encoding strategy is known as \textit{dense angle encoding}~\cite{LaRose2020RobustDE}. Employing this method, one qubit is utilized to encode two features serving as angles for two rotation gates $RY(x_i)$ or $RX(x_i)$ by using a phase gate after each rotation. It allows to encodes $N$ features in $N/2$ qubits. 
\begin{figure}[t]
\centering\includegraphics[width=0.48\textwidth]{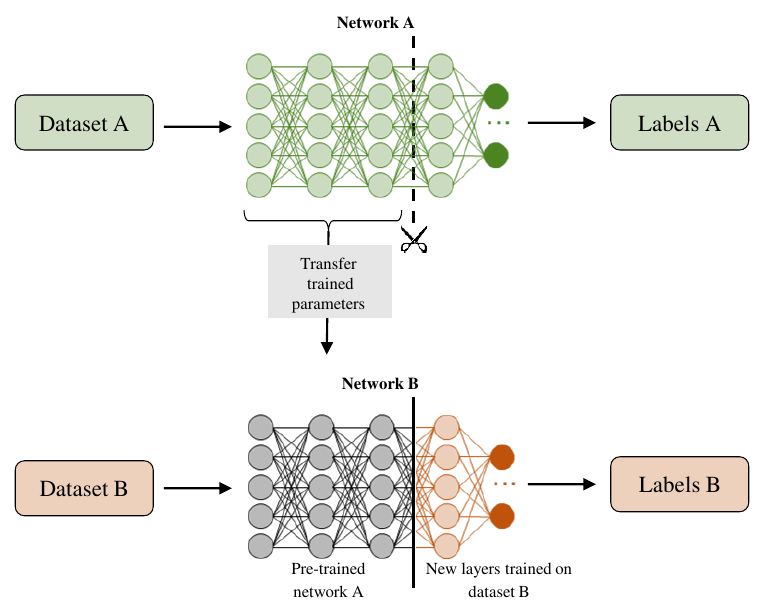}
    \caption {Overview of transfer learning architecture.
    The network A (green) is initially trained on the dataset A, and its weights are then transferred to the network B (orange). The last layer of A  is replaced in the network B by a new layer, which is trained on the dataset B. The remaining layers of the network A (grey) do not change during  training on the dataset B.}
    \label{fig:TransferLearning}
\end{figure}
\subsection{Quantum Machine Learning}
\begin{figure*}[t!]
\centering \includegraphics[width=1\textwidth]
{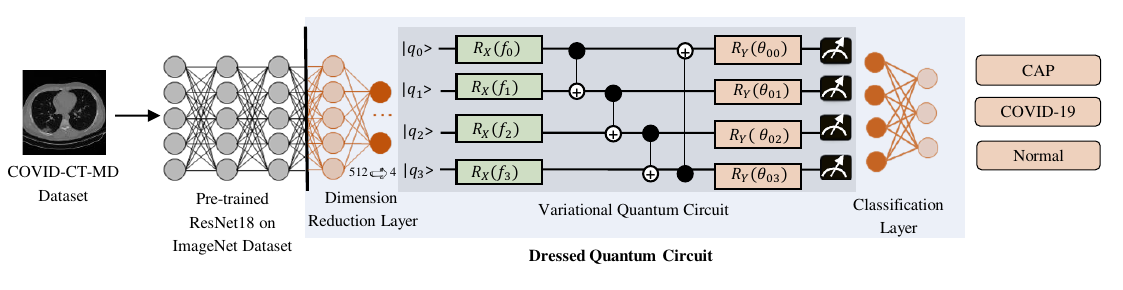}
    \caption {Overview of the experimental hybrid transfer learning architecture with dressed quantum circuit with depth 1. }
    \label{fig:DressedQuantumCircuit}
\end{figure*}
To overcome current limitations in QC, hybrid algorithms are a popular choice, in particular in QML, instead of using algorithms that exclusively run on quantum computers. Variational quantum algorithms (VQAs) are hybrid algorithms that operate iteratively quantum and classical computers, i.e., consist of a classical and a quantum part running on the respective machine. The core component of a VQA is the variational quantum circuit (VQC), a parameterized quantum circuit. 
For a classification problem, a VQC usually starts by embedding classical data, i.e., transforming classical input data to a quantum state. This is followed by repeating layers of rotation and entangling gates, where the rotation angles~$\theta$ are the parameters adjusted by an optimization algorithm running on a classical computer such that the cost function is minimized.
In the last step the qubits are measured and the measurement results can then be interpreted for a class prediction \cite{mitarai2018quantum,schuld2020circuit,cerezo2021variational}.
\subsection{Quantum Transfer Learning}
The main idea behind transfer learning is to reuse existing knowledge by combining a pre-trained network (A) with a separate (new) network (B) rather than training an entirely new network from scratch. The first network (A) is usually a large neural network trained on a generic task whose final layer is replaced by network B. The combined network is then applied to a related task for what network A was originally trained for. The architecture is displayed in Figure \ref{fig:TransferLearning}. 
If the weights of the pre-trained network A are frozen, only the weights of network B will be trained, i.e., updated. This process is known as \textit{feature extraction}, the original network is used as a feature extractor. The alternative approach where the weights of the original network A are also trained is known as \textit{fine tuning}.

Mari et al. proposed a hybrid classical-quantum transfer learning approach in \cite{mari2019transfer}. Their approach uses a classical neural network as network A and a dressed quantum circuit (DQC) as network B, however, they also discuss other approaches. They furthermore introduce their DQC, a VQC surrounded by a classical pre-processing and a classical post-processing layer.

The QTL approach is particularly suitable for image classification tasks as the classical neural network drastically reduces the number of features. 
The reduced feature set is then further processed by the quantum part of the algorithm.
\section{Methods}\label{sec:methods}

Our QTL approach is similar to that of Mari et al. \cite{mari2019transfer}. That is, we use ResNet18 as classical network, network A, while the second network, network B, is a quantum circuit. Note that we use a dressed quantum circuit as defined above in one approach, however, we also use a regular variational quantum circuit in a second approach. In this section, we discuss our QTL approaches in detail, i.e., we describe the overall architecture of our QTL variants. Moreover, we introduce the VQCs we utilize within the QTL networks.

\subsection{Dressed Quantum Circuit}
\begin{figure*}[t!]
\centering \includegraphics[width=1\textwidth]{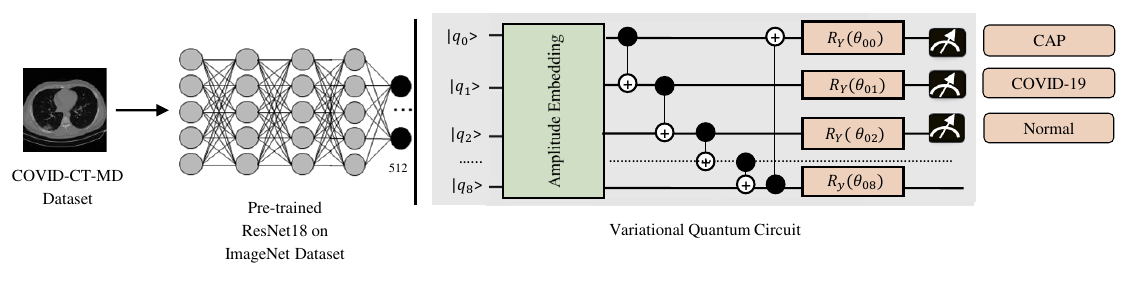}
    \caption {Overview of the experimental hybrid transfer learning architecture with variational quantum circuit only.}
\label{fig:StrictlySeparated}
\end{figure*}
Recall that a DQC is a variational quantum circuit surrounded by two classical fully connected layers, a pre- and a post-processing layer. The exact number of neurons and connections depends on the number of qubits in the quantum circuit. 
In our case, the classical pre-processing layer connects 512 ResNet18 output features to either 4 or 8 neurons (depending on the number of qubits) and thus serves as a dimension reduction layer while the post-processing layer connects either 4 or 8 neurons to either 2 or 3 neurons (depending on the number of classes) and hence acts as a classification layer. 
These additional classical layers greatly increase the number of parameters that are being trained, however, only for the classical layers as the number of trainable parameters for the quantum circuit depend on its depth and number of qubits. Furthermore, note that the parameters for the classical layers as well as for the VQC are trained simultaneously. While the pre-processing layer has the advantage of further reducing the dimensions of the features (from 512 to either 4 or 8 in the case of ResNet18), it is difficult to determine what impact each part (classical vs. quantum) have on the overall performance. Moreover, the classical part contains vastly more parameters than the quantum one. We will come back to this fact in the discussion of the results in the next chapter.

With the DQC, we use various circuit architectures and embedding methods, including angle embedding, where we embed one feature per qubit, and an angle embedding approach, where we embed two features in one qubit using a RX rotation for one feature and a RY rotation for another. The overall architecture of our QTL approach using a DQC is depicted in Fig. \ref{fig:DressedQuantumCircuit}. 

\subsection{Variational Quantum Circuit}
We also use a regular VQC for the quantum part in our QTL approach. That is, we do not include any classical pre- or post-processing layers. 
However, as described above, the output of ResNet18 second last layer is 512 features, too many to embed using angle encoding, since 512 qubits (or 256 qubits for the advanced approach using two rotations mentioned above) are required, far too many for simulators or quantum computers of the NISQ-era. 
We instead use amplitude embedding. With this embedding, only 9 qubits are required to embed all 512 features and thus a pre-processing layer is not required. The post-processing layer is also not necessary as we measure one qubit per class and apply \textit{Softmax} to get the prediction. The architecture of this QTL approach is shown in Fig. \ref{fig:StrictlySeparated}.

\subsection{Quantum Circuits}
We implemented various quantum circuits for our experiments. These circuits were deliberately kept "simple", i.e., they follow a straightforward pattern. More specifically, we first embed the features using an appropriate scheme (angle or amplitude embedding). This is followed by repeated layers of parameterized rotation and entangling gates. We used a "nearest neighbour" entangling method, i.e., adjacent qubits as well as the last and first qubit perform a CNOT gate. The number of repeating layers is determined by the "depth" parameter. The circuits are shown in Fig. \ref{fig:Circuits} in the appendix.

\section{Experimental Setup and Results}\label{sec:experimental_setup_and_results}
In this section, we discuss our experimental setup and present the results of the experiments. The main work focuses on the classification of COVID-19 from CT-scans, however, we also performed experiments on a benchmark dataset from the medical domain in order to evaluate the proposed approach on a related task. Before discussing the results, we introduce the dataset and the details of the configuration settings used in our experiments.

\subsection{Experiments Configuration}
Circuits were executed with 4 and 8 qubits, each with a depth of 1, 2 and 4. In order to facilitate the reproducibility of our results we ran all our experiments with seeds. The following evaluation and the related plots summarize the results over the parameters depth and seed for the specific number of qubits. 
The number of classical features embedded in the quantum circuit is dependent on the circuit architecture, as discussed in the previous section. Training was run for 45 epochs in the multi-class experiments on the COVID-19 dataset while 25 training epochs were run on the PneumoniaMNIST and OrganAMNIST dataset as well as the binary experiments on the COVID-19 dataset. All circuits were implemented using the PennyLane framework \cite{bergholm2018pennylane}. We use Adam~\cite{Kingma2014AdamAM} and the cross entropy loss to optimize parameters for both classical network and quantum circuit. 
To prevent overfitting, we use weight decay with the value of 0.01 in the optimizer. 
Furthermore, in all training setups we employed the same value 8 and 0.0001 for batch size and learning rate respectively.
Final model selection was based on the value of the validation AUROC score. 

\subsection{Data}
\setlength\tabcolsep{2.5pt}

To conduct our experiments, we use the COVID-CT-MD dataset~\cite{Afshar2021}. The dataset contains 512$\times$512 volumetric chest CT scans of 169 patients diagnosed positive for COVID-19 infection, 
60 patients with Community-Acquired Pneumonia (CAP) and 76 healthy patients.
All these three cases are collected from Babak Imaging Center in Tehran, Iran, and labeled by three
experienced radiologists in patient-level, slice-level, and lobe-level manners. 

In our experiments, we utilize slice-level labels only, whereby each slice is treated as an independent input for the classifier. 
After converting data from DICOM to PNG format, we produce a balanced multi-class splitting. 
To achieve this, we begin by filtering out slices for each COVID-19 and CAP patient, since the lung changes are not as obvious on all series. 
After this process is finished, 1178 CAP examples (the smallest class) remain. Therefore, we left the same number of images for COVID-19 and healthy cases and split data into train~(831 images), validation~(184 images) and test~(163 images) sets. Statistics of patients and data examples is depicted in Table~\ref{tab:COVIDDataMutliclass}.  
For the balanced binary splitting of the dataset, we removed the "CAP" class while for the remaining classes "COVID-19" and "Normal" we retained the same number of images to facilitate more accurate comparison of results.
Another important point to consider is that the CT images from the same patient are highly correlated. Therefore, the images from a particular patient can only be included in one subset. Once any CT image of the patient is assigned to the train, the validation, or the test subset, the rest of the slices can be obtained in the same subset only.  

\begin{table}
\centering
\caption {Statistics of patients and images in the multi-class version of the COVID-CT-MD dataset.}
\begin{tabular}{|c|cc|cc|cc|}
\hline
\textbf{Class}    & \multicolumn{2}{c|}{\textbf{Train set $\sim$70\%}}       & \multicolumn{2}{c|}{\textbf{Validation set $\sim$15\%}}     & \multicolumn{2}{c|}{\textbf{Test set $\sim$15\%)}}           \\ \hline
         & \multicolumn{1}{c|}{Patients} & Images & \multicolumn{1}{c|}{Patients} & Images & \multicolumn{1}{c|}{Patients} & Images \\ \hline
Cap      & \multicolumn{1}{c|}{16}              & 831           & \multicolumn{1}{c|}{4}               & 184           & \multicolumn{1}{c|}{5}               & 163           \\ \hline
COVID-19 & \multicolumn{1}{c|}{29}              & 831           & \multicolumn{1}{c|}{15}              & 184           & \multicolumn{1}{c|}{9}               & 163           \\ \hline
Normal   & \multicolumn{1}{c|}{98}              & 831           & \multicolumn{1}{c|}{31}              & 184           & \multicolumn{1}{c|}{25}              & 163           \\ \hline
\end{tabular}
\label{tab:COVIDDataMutliclass}
\end{table}

\subsection{COVID-19 Experiments}
We ran a range of experiments on the dataset described above and applied it to multi-class classification using the three classes "Normal", "CAP" and "COVID-19" and a binary classification for the classes "Normal" vs. "COVID-19". 

\begin{figure}[t]
  \centering
  \begin{tabular}{ c @{\hspace{0pt}} c }
    \includegraphics[width=1\linewidth, height=9cm] {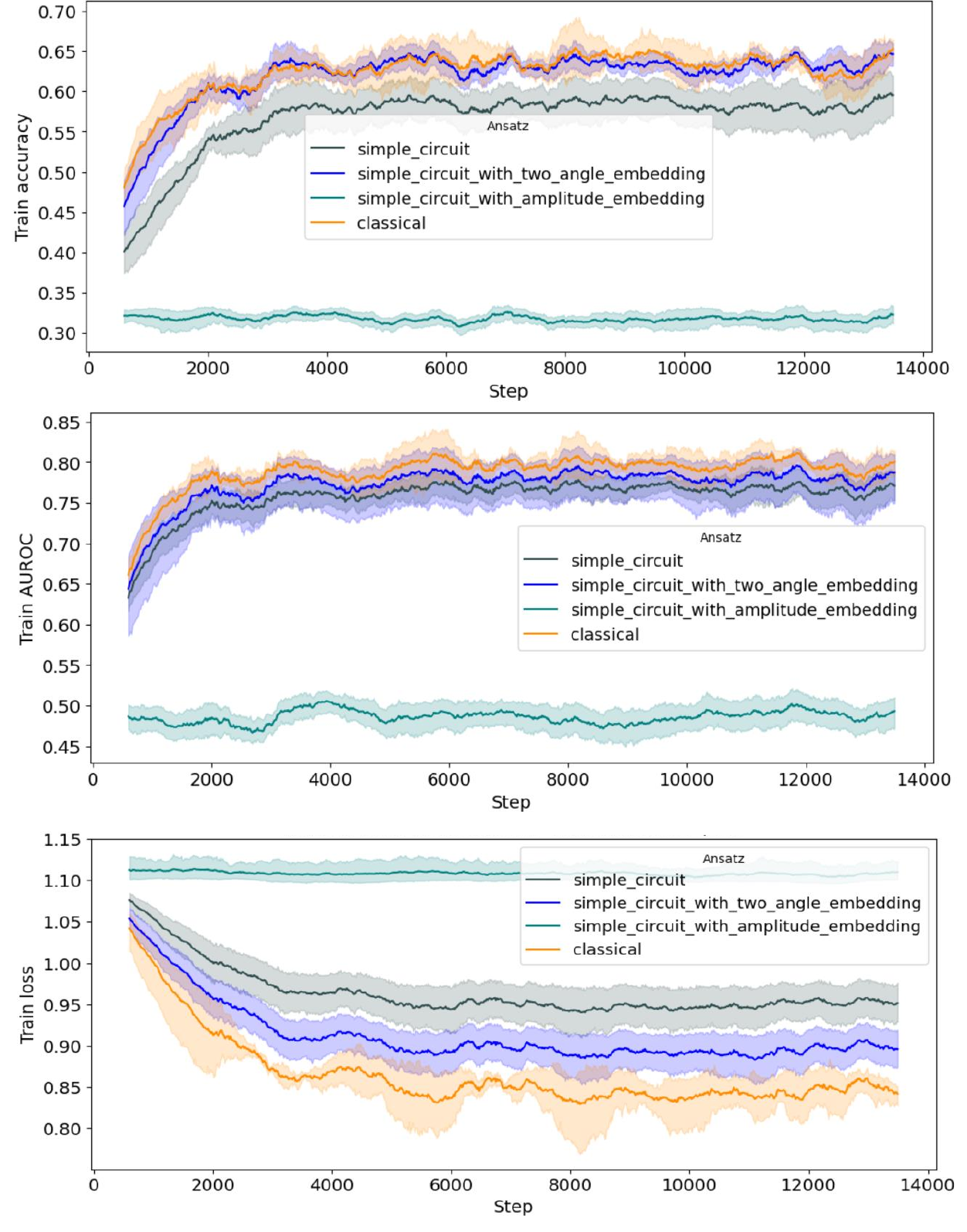} \\
     \footnotesize (a) Train metrics on multi-class COVID-CT-MD 
       \\
       \includegraphics[width=1\linewidth]{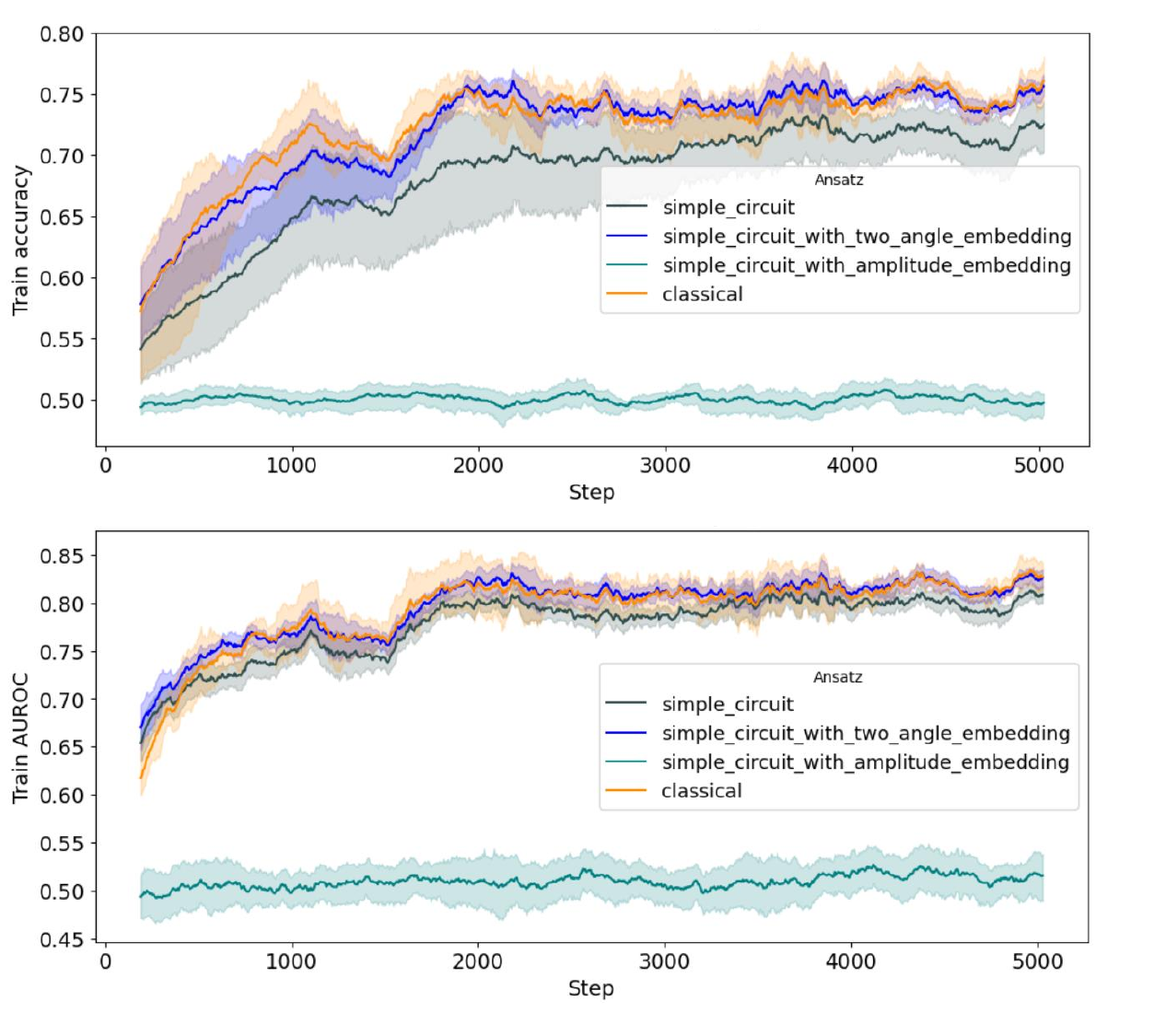} \\
      \footnotesize (b) Train metrics on binary COVID-CT-MD
  \end{tabular}
  \caption{Results on multi-class and binary COVID-CT-MD: 8~qubits in DQC approaches, 9 qubits in amplitude embedding approach and 8 hidden units in the classical network.}
  \label{fig:TrainCOVID}
\end{figure}

Training performance on the multi-class problem is depicted in Fig. \ref{fig:TrainCOVID}a. While the classical and the DQC approaches perform similar, all achieving an accuracy roughly in the range 0.55 and 0.60, the approach where only the quantum part is trained, i.e., where we use amplitude embedding rather than a quantum circuit with classical pre- and post-processing layers, does not improve over training.  
When 8 qubits are used in the DQC, the "simple circuit" that embeds 2 features into one qubit using 2 different angles performs similar to the classical approach. Note that while the approach uses more qubits, the number of classical features is also increased, as described in the previous section. 
In the middle plot of the figure, the AUROC value achieved over training is shown.
The bottom plot of the figure displays the loss values during the training phase. While loss values of the DQC and classical approaches decrease as expected, the loss of the circuit with amplitude embedding remains almost the same over training.

We furthermore evaluated the model on a binary problem, i.e., classifying CT-scans into either "COVID-19" or "Normal", the results can be seen in Fig. \ref{fig:TrainCOVID}b. In the binary setting, all models achieve better training accuracy than in the multi-class setting.
The DQC with two angle embedding and classical network reach almost the same train accuracy, roughly within the range 0.7-0.75, and perform slightly better than the DQC with a simple circuit.
However, the AUROC values achieved by the two DQCs and classical model are almost identical, roughly in the range 0.80 - 0.85. 
In this setting, the amplitude embedding circuit still does not perform well, especially when compared to the other approaches. Closer inspection revealed that in this case the model tends to predict only one class (either "COVID-19" or "Normal") while disregarding the other, even though the data is balanced.  

\begin{figure}[ht]
  \centering
  \begin{tabular}{ c @{\hspace{0pt}} c }\includegraphics[width=1\linewidth, height=7cm] {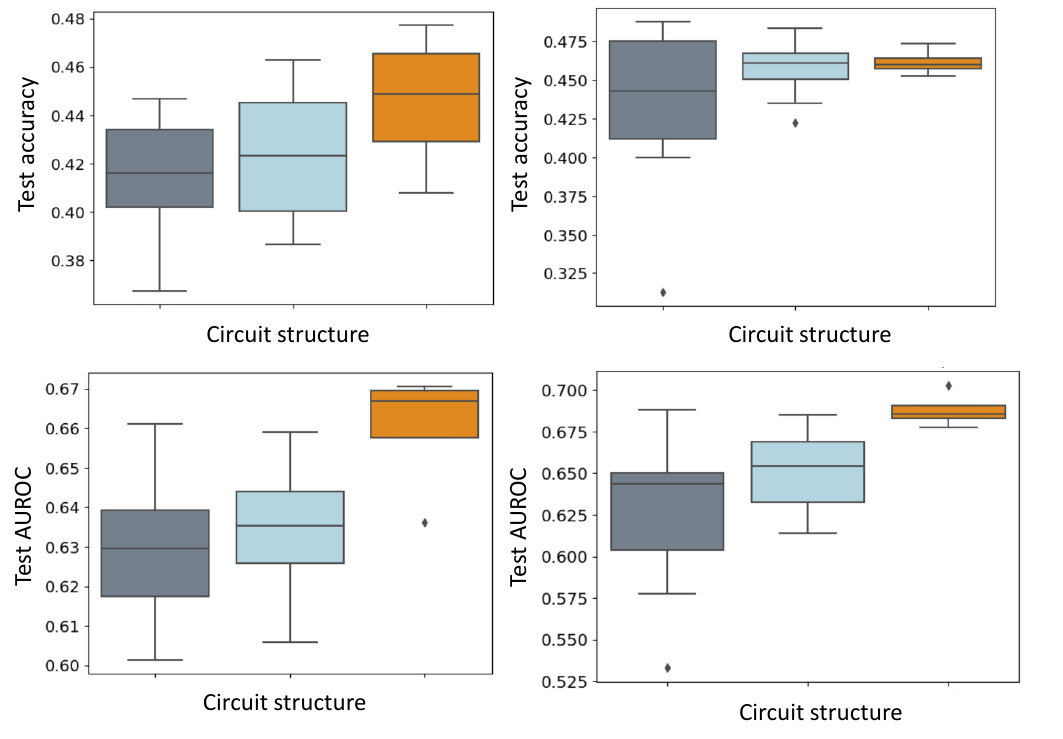} \\
    \footnotesize (a) Test metrics on multi-class COVID-CT-MD \\ \\\includegraphics[width=1\linewidth, height=7cm]{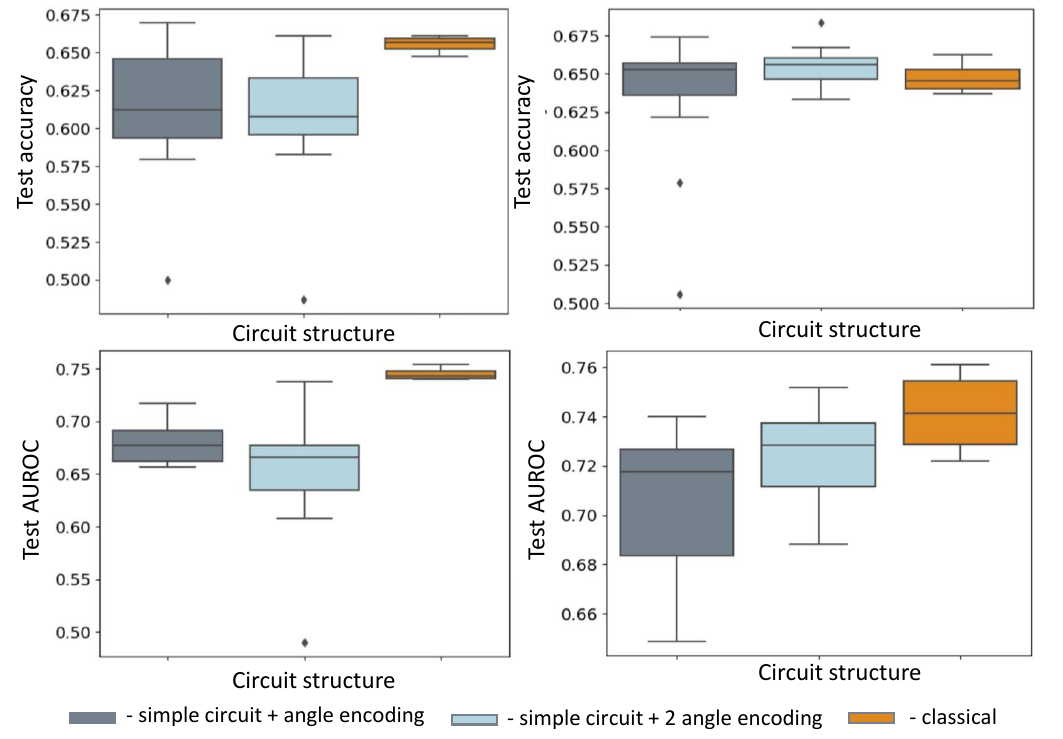} \\
      \footnotesize (b) Test metrics on binary COVID-CT-MD
  \end{tabular}
  \caption{Results on multi-class and binary COVID-CT-MD.  \textit{Left}: DQC 4 qubits, 4 hidden units in classical. \textit{Right}: DQC 8 qubits, 8 hidden units in classical. \textit{Grey}: DQC with a "simple circuit" and a standard angle encoding. \textit{Blue}: DQC with a "simple circuit" and an angle encoding that embeds 2 features into one
qubit using 2 different angles. \textit{Orange}: Classical approach. }
\label{fig:COVIDest}
\end{figure}

\begin{figure}[bt]
\centering  \includegraphics[width=0.5\textwidth]{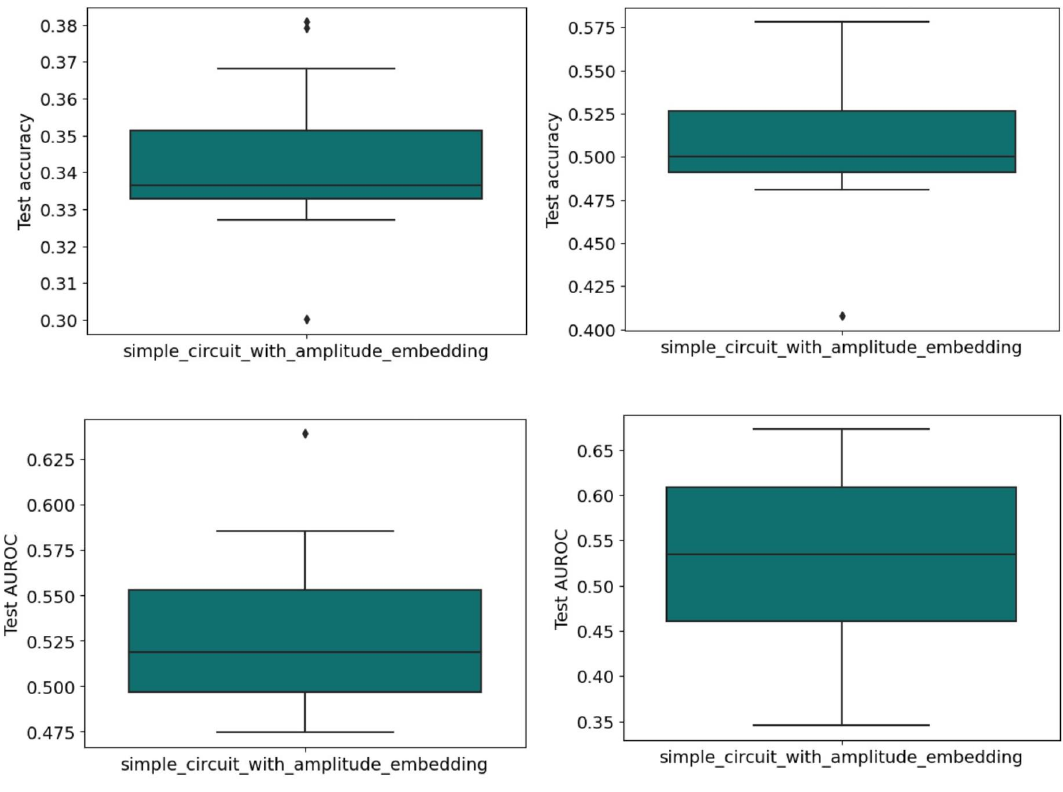}
    \caption {Results on multi-class COVID-CT-MD (left) and binary COVID-CT-MD (right) test data: QTL approach with amplitude embedding.}
    \label{fig:COVIDAETest}
\end{figure}

Results on the test subset are shown in Fig. \ref{fig:COVIDest}a (multi-class) and Fig. \ref{fig:COVIDest}b (binary). Results on the test data for our QTL approach using amplitude embedding can be seen in Fig.~\ref{fig:COVIDAETest}.  


\begin{figure}[tb]
\centering \includegraphics[width=0.5\textwidth]{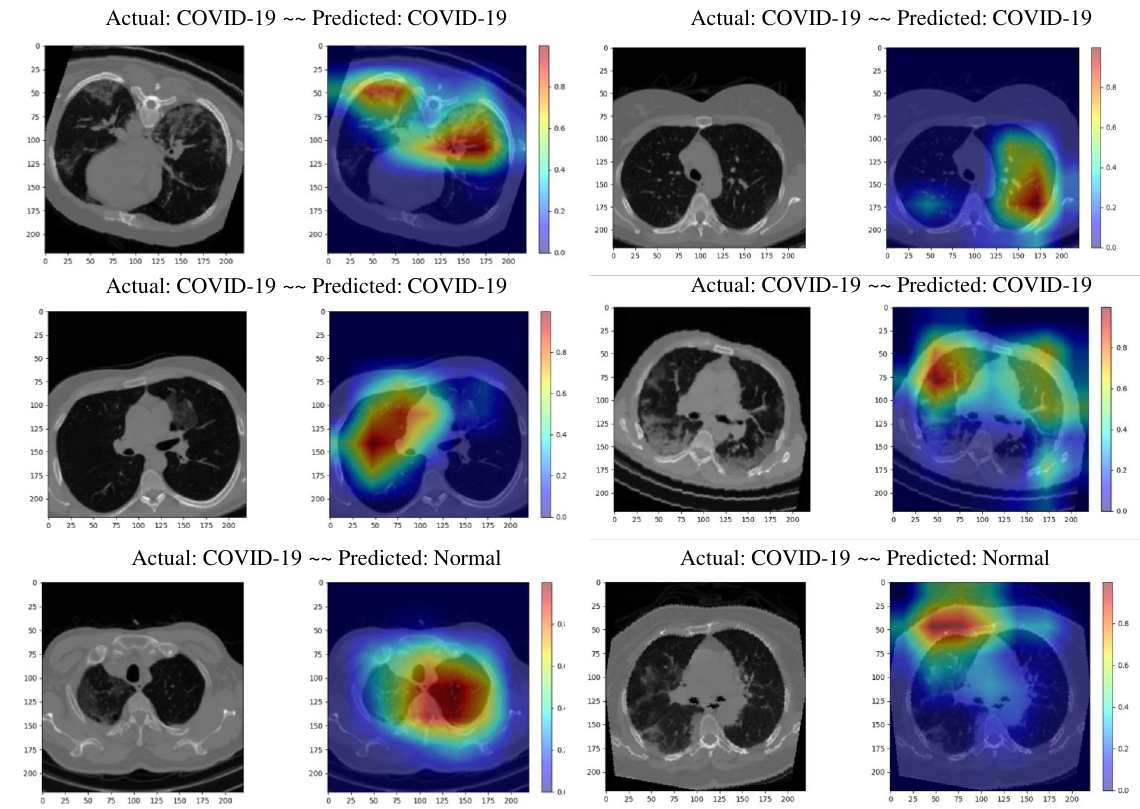}
\caption {Pathology localization through the DQC model on a binary version of the COVID-CT-MD dataset. Areas highlighted in red are most important for the model decision in particular cases. The top row depicts examples, where the model classifies the input correctly and focuses on areas with pathologies. The middle row shows examples, where the model also gives a correct prediction but focuses on areas that are considered irrelevant by experts. The bottom row displays examples of false classified images.}
\label{fig:Heatmaps}
\end{figure}

The explainability component is a crucial part of any machine learning application in the medical context.  
To visualize model decisions, we apply Grad-CAM++~\cite{GradCAMPlusPlus} on the hybrid network. To provide an explanation of which areas of the input image influenced the model prediction the most, GradCAM++ makes use of extracting positive gradients of the final convolutional layer feature maps and weighting it w.r.t. a specific class score.  GradCAM++ works with any CNN architecture without any retraining but needs the final score for the particular class to be a differentiable function of the last convolution layer activation maps. The result of the GradCAM++ technique is a heat map (or class activation map), which highlights the important parts of the input image. Fig.~\ref{fig:Heatmaps} shows heat maps produced by a DQC model trained on the binary version of the COVID-CT-MD dataset. 

\subsection{Application to other datasets}
In order to further evaluate our approach, we ran experiments on other benchmarking datasets from the medical domain. For this we used the binary dataset PneumoniaMNIST and multi-class dataset OrganAMNIST from MedMNIST v2~\cite{medmnistv2}. These 28$\times$28 images are substantially smaller than the CT images used in previous experiments. Note that we ran the following experiments with 4 different seeds. Table~\ref{tab:MedMNISTStatistics} shows the statistics on the data and splits used for these experiments.
\begin{table}[t!]
\centering
\caption{Dataset sizes of PneumoniaMNIST and OrganaMNIST}
\begin{tabular}{|cccc|}
\hline
\multicolumn{4}{|c|}{\textbf{PneumoniaMNIST}}                                                                                                     \\ \hline
\multicolumn{1}{|c|}{\textbf{Class}} & \multicolumn{1}{c|}{\hspace{0.3cm}\textbf{Train set} \hspace{0.3cm}\ } & \multicolumn{1}{c|}{\hspace{0.3cm}\textbf{Validation set}\hspace{0.5cm}\ } & \textbf{\hspace{0.3cm}Test set}\hspace{0.5cm} \\ \hline
\multicolumn{1}{|c|}{\hspace{0.3cm}Normal\hspace{0.5cm}}         & \multicolumn{1}{c|}{1214}               & \multicolumn{1}{c|}{135}                     & 234               \\ \hline
\multicolumn{1}{|c|}{\hspace{0.3cm}Pneumonia\hspace{0.5cm}\ }      & \multicolumn{1}{c|}{3494}               & \multicolumn{1}{c|}{389}                     & 390               \\ \hline
\multicolumn{4}{|c|}{\textbf{OrganaMNIST}}                                                                                                        \\ \hline
\multicolumn{1}{|c|}{Bladder}        & \multicolumn{1}{c|}{1956}               & \multicolumn{1}{c|}{321}                     & 1036              \\ \hline
\multicolumn{1}{|c|}{Femur-left}     & \multicolumn{1}{c|}{1408}               & \multicolumn{1}{c|}{233}                     & 784               \\ \hline
\multicolumn{1}{|c|}{Heart}          & \multicolumn{1}{c|}{1474}               & \multicolumn{1}{c|}{392}                     & 785               \\ \hline
\end{tabular}
\label{tab:MedMNISTStatistics}
\end{table}

\subsubsection{PneumoniaMNIST Experiments}
\begin{figure}[t]
  \centering
  \begin{tabular}{ c @{\hspace{0pt}} c }
\includegraphics[width=0.9\linewidth] {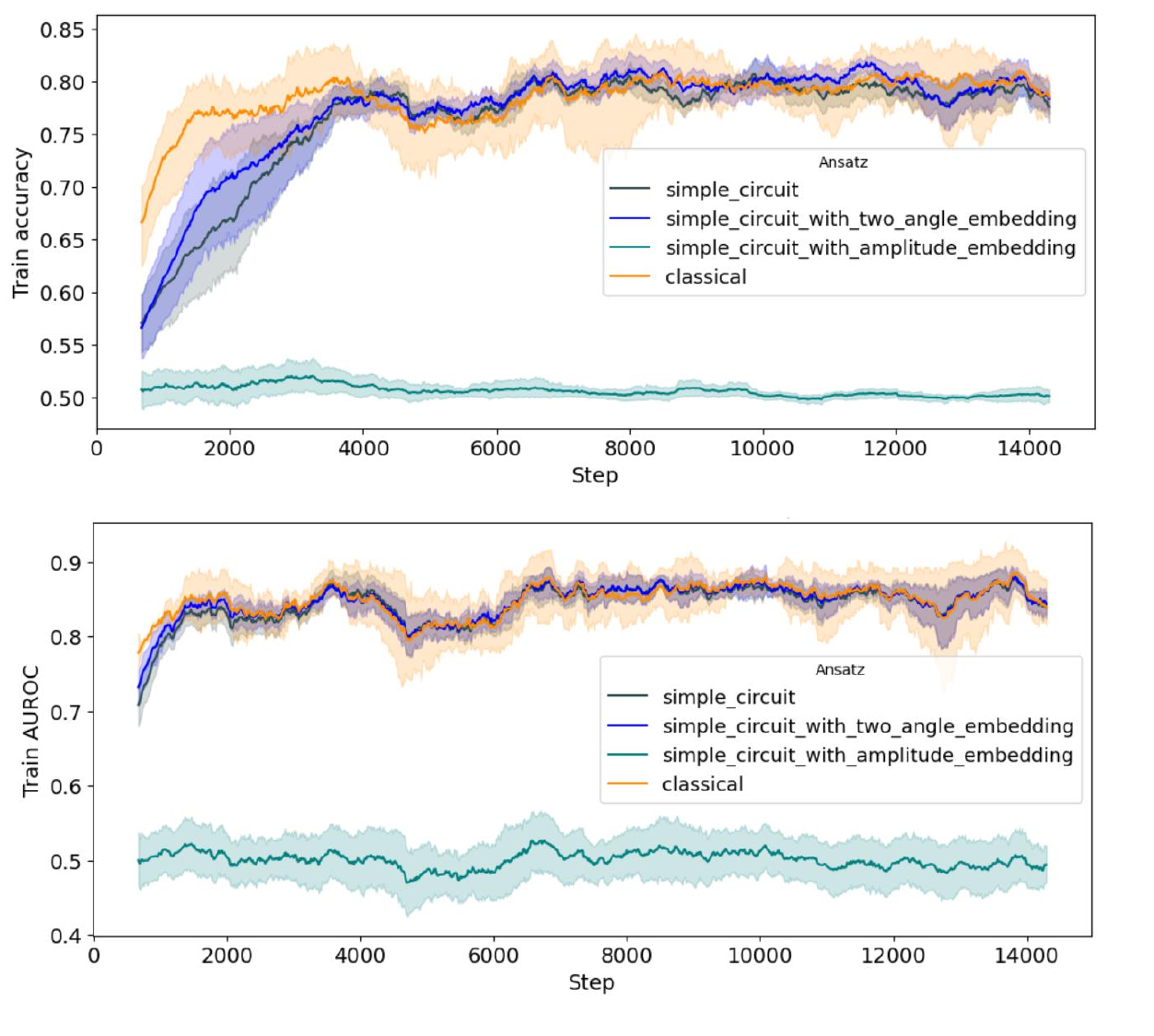} \\
    \footnotesize (a) Train metrics on PneumoniaMNIST \\
    \\
     \includegraphics[width=0.9\linewidth]{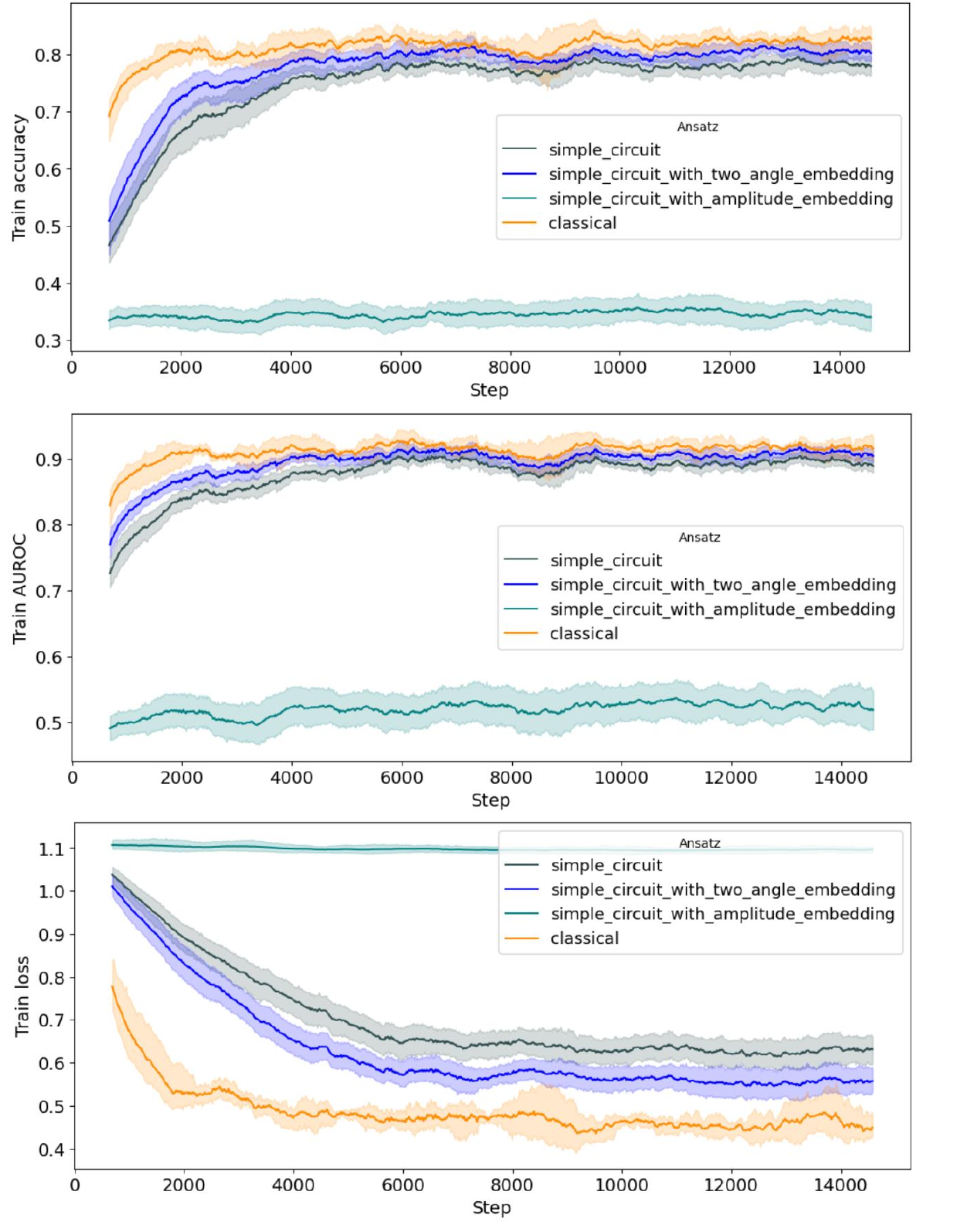}\\
      \footnotesize (b) Train metrics on OrganAMNIST
  \end{tabular}
  \caption{Results on MedMNIST train data: DQC 8 qubits, 8 hidden units in classical, 9 qubits amplitude embedding.}
\label{fig:image2}
\end{figure}

Training accuracy as well as AUROC on the binary pneumonia classification problem is shown in Fig. \ref{fig:image2}a. For both metrics, the classical model as well as both DQCs achieved similar results, which are relatively good when compared to the COVID-19 experiments. The amplitude embedding circuit also performs far worse in this setup than the other approaches. Furthermore, the AUROC value does not improve much over the entire training process. 
Various factors may be responsible for these results, and we will discuss some issues in the next section.

\begin{figure}[tb]
  \centering
  \begin{tabular}{ c @{\hspace{20pt}} c }
    \includegraphics[width=0.9\linewidth] {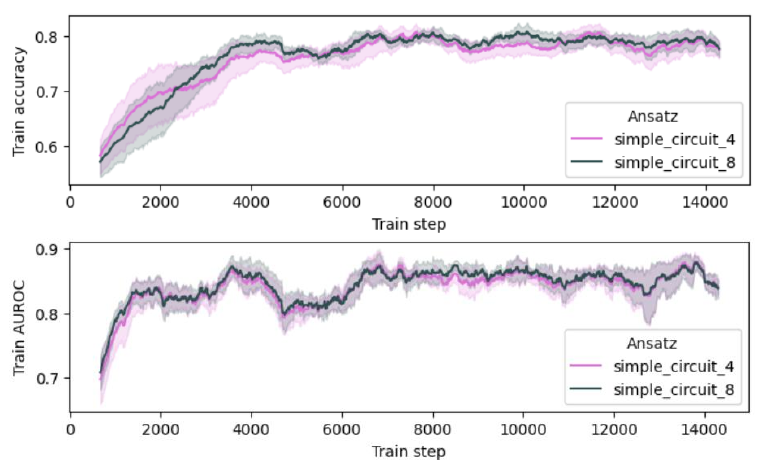} \\
    \footnotesize (a) DQC with a simple circuit and angle embedding \\ \\
      \includegraphics[width=0.9\linewidth]{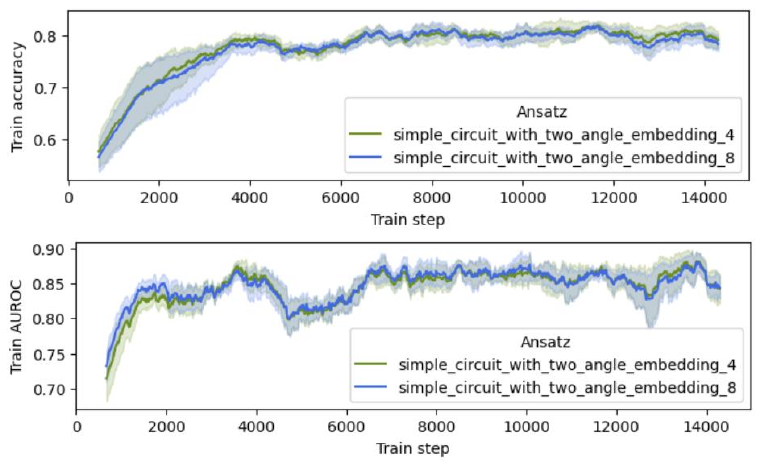} \\
      \footnotesize (b) DQC with a simple circuit and two angle embedding
  \end{tabular}
\caption{Performance comparison of DQC approaches with 4  vs. 8 qubits on PneumoniaMNIST.}
\label{fig:PneumoniaQubits}
\end{figure}

\begin{figure}[ht]
  \centering
  \begin{tabular}{ c @{\hspace{0pt}} c }
\includegraphics[width=0.95\linewidth] {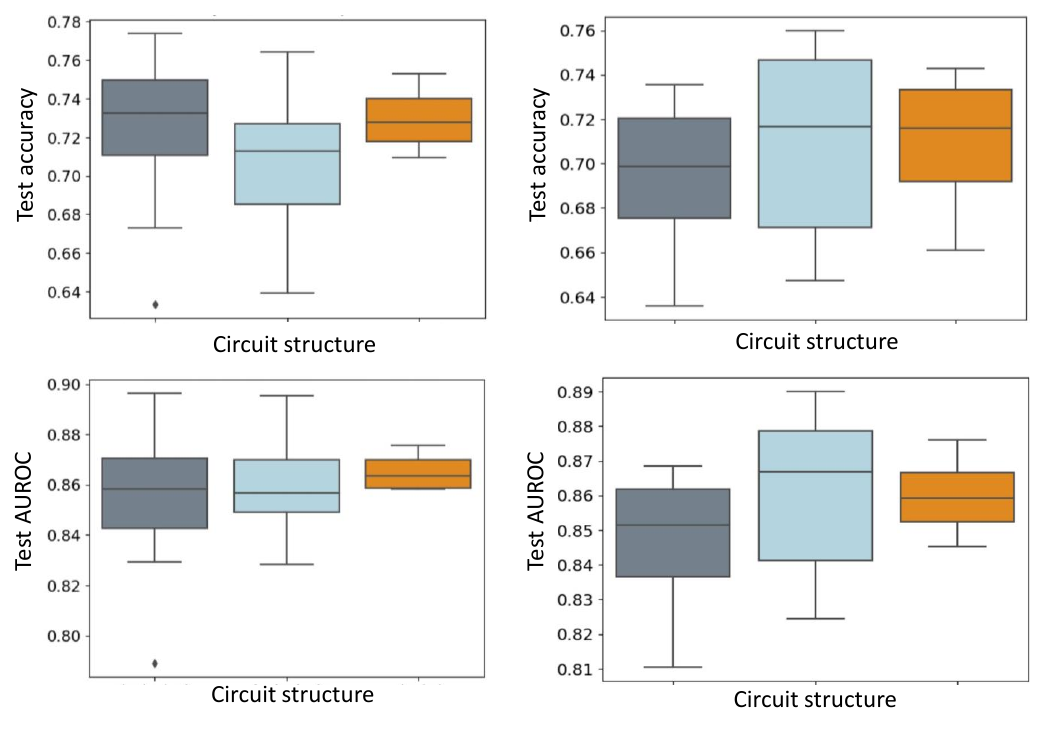} \\
    \footnotesize (a) Test metrics on PneumoniaMNIST \\ \\
\includegraphics[width=0.95\linewidth]{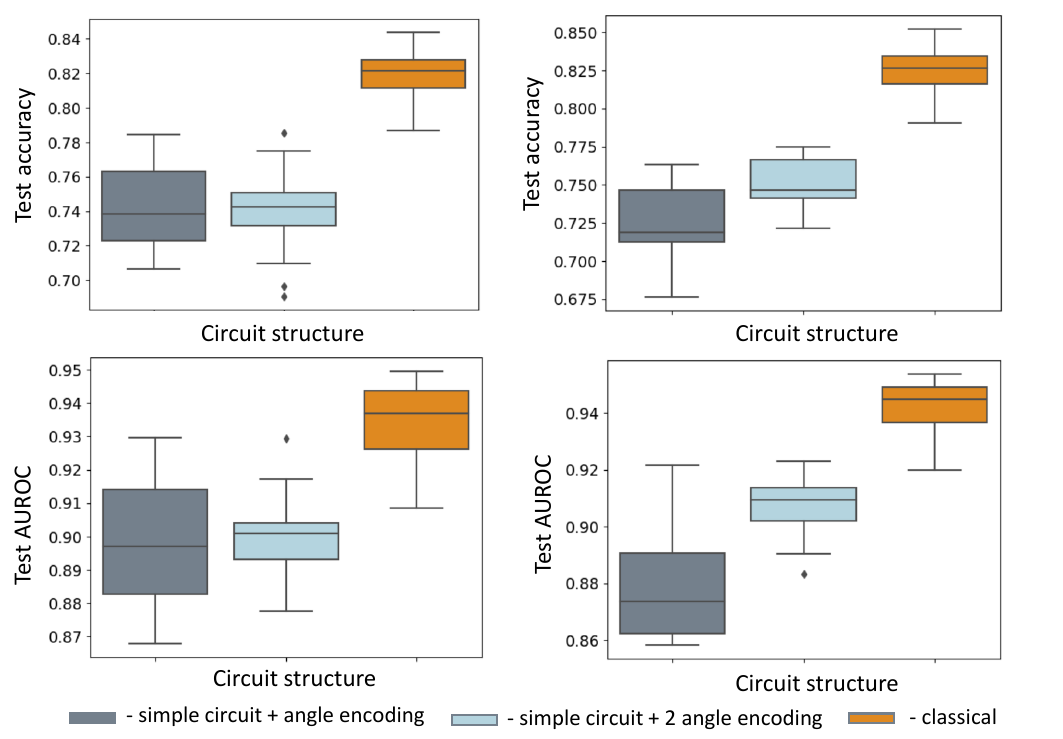} \\
      \footnotesize (b) Test metrics on OrganAMNIST
  \end{tabular}
  \caption{Results on MedMNIST: \textit{Left}: DQC 4 qubits, 4 hidden units in classical. \textit{Right}: DQC 8 qubits, 8 hidden units in classical.}
\label{fig:image3}
\end{figure}

We evaluated our quantum circuits with 4 and 8 qubits, a comparison on the PneumoniaMNIST dataset can be seen in Fig. \ref{fig:PneumoniaQubits}. While the accuracy values vary to some degree over training, the end results are mostly the same. The AUROC scores remain fairly close throughout the training process.
Test accuracy and AUROC score from this experiment can be seen in Fig.~\ref{fig:image3}a.

\subsubsection{OrganAMNIST Experiments}
While the OrganAMNIST dataset originally has 10 classes, we adjusted it to only include 3 classes ("bladder", "femur-left" and "heart"). Training performance on this dataset is shown in Fig. \ref{fig:image2}b. On this dataset, the model achieves higher accuracy and AUROC scores than on the COVID-19 multi-class task, except the quantum circuit using amplitude embedding, where the results are similar. Test results are depicted in Fig~\ref{fig:image3}b.

\section{Discussion}
\label{sec:discussion}
In this work, our main focus was to evaluate different QTL approaches on an image classification task from the medical domain, namely classifying CT-scans of the lung into one of three classes. However, we also evaluated our model on other datasets for related problems from the medical domain. While the applied models work fairly well on the PneumoniaMNIST and OrganAMNIST datasets, binary and multi-class tasks respectively, it does not perform that well on the COVID-CT-MD dataset, especially on the test data. There could be several reasons for this. The COVID-19 dataset is relatively small with an unequal amount of patients for each class (although the number of images was balanced), which could affect training and hinder generalization, the PneumoniaMNIST and OrganAMNIST datasets on the other hand contain vastly more training data. 
Using circuits with more qubits and layers (and thus more parameters) may also yield better results as this may increase the models capacity, however, more investigation in this regard is required.

An important note to stress, though, is the fact that the DQC is not entirely "quantum", so to speak, as it is surrounded by a classical pre- and a post-processing layer, each with classical parameters that are being simultaneously optimized as the parameters of the quantum part. It is thus difficult to determine what part is ultimately responsible for the overall performance. Also note that the number of classical parameters vastly outnumber the parameters used in the quantum circuit.

While our second approach uses amplitude embedding to embed all 512 features resulting from ResNet18, it does not contain any classical layers that are trained further. That is, only the angles of the rotation gates in the quantum circuit are being optimized and trained on the respective dataset. It is then furthermore crucial to point out that we ran experiments with depth 1, 2 and 4, and thus only up to 36 parameters in this architecture were trained on the medical dataset, where the number of parameters depends on the depth of the circuit. Increasing the number of layers and hence trainable parameters may result in a better performance as this can increase the models expressibility.

\section{Conclusion}
\label{sec:conclusion}
We applied different QTL variants to the task of classifying CT-scans of the lung into either "Normal", "CAP" or "COVID-19". As CT-scans are fairly large and because the capabilities of current QC-hardware are not sufficient to be able to run large quantum circuits, we chose a hybrid approach. In this setting, a classical pre-trained network such as ResNet18 acts as a feature extractor and reduces the dimensions such that these reduced number of features can be further processed by a quantum circuit. The results show that it is a difficult task to classify large CT-scans in a multi-class setting as well as in the binary case, albeit the results are slightly better in the latter. Training on a larger dataset or increasing the depth of the quantum circuits may result in a better performance on the COVID-19 classification task. To further evaluate our approach, we applied it to related tasks from the medical domain, the classification of pneumonia and organs, these datasets being binary and multi-class respectively. Our approach yielded better results on these datasets, although these images are also smaller and the datasets contain more images.
Further work is required to unleash the potential of applying QC to problems of this magnitude, and whether this will be the case in the NISQ-era remains to be seen. Until then, though, several research avenues may be explored. Different circuit architectures as well as their trainability and ability to avoid barren plateaus could be evaluated in this context. In addition to this, efficient image specific embedding techniques should be considered. Furthermore, other means of feature reduction should be investigated, in particular such methods where the classical and quantum parts are strictly separated, i.e., are trained separately. As for medical image classification, image segmentation prior to classification may also be a worthwhile endeavour. 


\bibliographystyle{IEEEtran}
\bibliography{IEEEabrv, bibliography}

\begin{thebibliography}{10}
\providecommand{\url}[1]{#1}
\csname url@samestyle\endcsname
\providecommand{\newblock}{\relax}
\providecommand{\bibinfo}[2]{#2}
\providecommand{\BIBentrySTDinterwordspacing}{\spaceskip=0pt\relax}
\providecommand{\BIBentryALTinterwordstretchfactor}{4}
\providecommand{\BIBentryALTinterwordspacing}{\spaceskip=\fontdimen2\font plus
\BIBentryALTinterwordstretchfactor\fontdimen3\font minus
  \fontdimen4\font\relax}
\providecommand{\BIBforeignlanguage}[2]{{%
\expandafter\ifx\csname l@#1\endcsname\relax
\typeout{** WARNING: IEEEtran.bst: No hyphenation pattern has been}%
\typeout{** loaded for the language `#1'. Using the pattern for}%
\typeout{** the default language instead.}%
\else
\language=\csname l@#1\endcsname
\fi
#2}}
\providecommand{\BIBdecl}{\relax}
\BIBdecl

\bibitem{herman2022survey}
D.~Herman, C.~Googin, X.~Liu, A.~Galda, I.~Safro, Y.~Sun, M.~Pistoia, and
  Y.~Alexeev, ``A survey of quantum computing for finance,'' \emph{arXiv
  preprint arXiv:2201.02773}, 2022.

\bibitem{egger2020quantum}
D.~J. Egger, C.~Gambella, J.~Marecek, S.~McFaddin, M.~Mevissen, R.~Raymond,
  A.~Simonetto, S.~Woerner, and E.~Yndurain, ``Quantum computing for finance:
  State-of-the-art and future prospects,'' \emph{IEEE Transactions on Quantum
  Engineering}, vol.~1, pp. 1--24, 2020.

\bibitem{kassal2011simulating}
I.~Kassal, J.~D. Whitfield, A.~Perdomo-Ortiz, M.-H. Yung, and A.~Aspuru-Guzik,
  ``Simulating chemistry using quantum computers,'' \emph{Annual review of
  physical chemistry}, vol.~62, pp. 185--207, 2011.

\bibitem{cao2019quantum}
Y.~Cao, J.~Romero, J.~P. Olson, M.~Degroote, P.~D. Johnson, M.~Kieferov{\'a},
  I.~D. Kivlichan, T.~Menke, B.~Peropadre, N.~P. Sawaya \emph{et~al.},
  ``Quantum chemistry in the age of quantum computing,'' \emph{Chemical
  reviews}, vol. 119, no.~19, pp. 10\,856--10\,915, 2019.

\bibitem{brown2010using}
K.~L. Brown, W.~J. Munro, and V.~M. Kendon, ``Using quantum computers for
  quantum simulation,'' \emph{Entropy}, vol.~12, no.~11, pp. 2268--2307, 2010.

\bibitem{schuld2015introduction}
M.~Schuld, I.~Sinayskiy, and F.~Petruccione, ``An introduction to quantum
  machine learning,'' \emph{Contemporary Physics}, vol.~56, no.~2, pp.
  172--185, 2015.

\bibitem{biamonte2017quantum}
J.~Biamonte, P.~Wittek, N.~Pancotti, P.~Rebentrost, N.~Wiebe, and S.~Lloyd,
  ``Quantum machine learning,'' \emph{Nature}, vol. 549, no. 7671, pp.
  195--202, 2017.

\bibitem{farhi2014quantum}
E.~Farhi, J.~Goldstone, and S.~Gutmann, ``A quantum approximate optimization
  algorithm,'' \emph{arXiv preprint arXiv:1411.4028}, 2014.

\bibitem{preskill2018quantum}
J.~Preskill, ``Quantum computing in the nisq era and beyond,'' \emph{Quantum},
  vol.~2, p.~79, 2018.

\bibitem{shor1999polynomial}
P.~W. Shor, ``Polynomial-time algorithms for prime factorization and discrete
  logarithms on a quantum computer,'' \emph{SIAM review}, vol.~41, no.~2, pp.
  303--332, 1999.

\bibitem{grover1996fast}
L.~K. Grover, ``A fast quantum mechanical algorithm for database search,'' in
  \emph{Proceedings of the twenty-eighth annual ACM symposium on Theory of
  computing}, 1996, pp. 212--219.

\bibitem{mari2019transfer}
A.~Mari, T.~R. Bromley, J.~Izaac, M.~Schuld, and N.~Killoran, ``Transfer
  learning in hybrid classical-quantum neural networks,'' \emph{arXiv preprint
  arXiv:1912.08278}, 2019.

\bibitem{icaart23}
P.~Altmann., L.~Sünkel., J.~Stein., T.~Müller., C.~Roch., and
  C.~Linnhoff{-}Popien., ``Sequent: Towards traceable quantum machine learning
  using sequential quantum enhanced training,'' in \emph{Proceedings of the
  15th International Conference on Agents and Artificial Intelligence - Volume
  3: ICAART,}, INSTICC.\hskip 1em plus 0.5em minus 0.4em\relax SciTePress,
  2023, pp. 744--751.

\bibitem{pramanik2022quantum}
S.~Pramanik, M.~G. Chandra, C.~Sridhar, A.~Kulkarni, P.~Sahoo, C.~D. Vishwa,
  H.~Sharma, V.~Navelkar, S.~Poojary, P.~Shah \emph{et~al.}, ``A
  quantum-classical hybrid method for image classification and segmentation,''
  in \emph{2022 IEEE/ACM 7th Symposium on Edge Computing (SEC)}.\hskip 1em plus
  0.5em minus 0.4em\relax IEEE, 2022, pp. 450--455.

\bibitem{otgonbaatar2022quantum}
S.~Otgonbaatar, G.~Schwarz, M.~Datcu, and D.~Kranzlmüller, ``Quantum transfer
  learning for real-world, small, and high-dimensional datasets,'' 2022.

\bibitem{azevedo2022quantum}
V.~Azevedo, C.~Silva, and I.~Dutra, ``Quantum transfer learning for breast
  cancer detection,'' \emph{Quantum Machine Intelligence}, vol.~4, no.~1, p.~5,
  2022.

\bibitem{majumdar2023histopathological}
R.~Majumdar, B.~Baral, B.~Bhalgamiya, and T.~D. Roy, ``Histopathological cancer
  detection using hybrid quantum computing,'' \emph{arXiv preprint
  arXiv:2302.04633}, 2023.

\bibitem{landman2022quantum}
J.~Landman, N.~Mathur, Y.~Y. Li, M.~Strahm, S.~Kazdaghli, A.~Prakash, and
  I.~Kerenidis, ``Quantum methods for neural networks and application to
  medical image classification,'' \emph{Quantum}, vol.~6, p. 881, 2022.

\bibitem{umer2022integrated}
M.~J. Umer, J.~Amin, M.~Sharif, M.~A. Anjum, F.~Azam, and J.~H. Shah, ``An
  integrated framework for covid-19 classification based on classical and
  quantum transfer learning from a chest radiograph,'' \emph{Concurrency and
  Computation: Practice and Experience}, vol.~34, no.~20, p. e6434, 2022.

\bibitem{Tahamtan2020}
A.~Tahamtan and A.~Ardebili, ``Real-time rt-pcr in covid-19 detection: issues
  affecting the results,'' \emph{Expert review of molecular diagnostics},
  vol.~20, no.~5, pp. 453--454, 2020.

\bibitem{Yang2020}
R.~Yang, X.~Li, H.~Liu, Y.~Zhen, X.~Zhang, Q.~X. Xiong, Y.~Luo, C.~Gao, and
  W.~Zeng, ``Chest ct severity score: An imaging tool for assessing severe
  covid-19,'' \emph{Radiology Cardiothoracic Imaging}, vol.~2, no.~2, 2020.

\bibitem{Fang2020}
Y.~Fang, H.~Zhang, J.~Xie, M.~Lin, L.~Ying, P.~Pang, and W.~Ji, ``Sensitivity
  of chest ct for covid-19: Comparison to rt-pcr,'' \emph{Radiology}, vol. 296,
  no.~2, 2020.

\bibitem{Ye2020}
Z.~Ye, Y.~Zhang, Z.~Huang, and B.~Song, ``Chest ct manifestations of new
  coronavirus disease 2019 (covid-19): a pictorial review,'' \emph{European
  Radiology}, vol.~30, pp. 4381--4389, 2020.

\bibitem{Sahu2020}
K.~Sahu, A.~Lal, and A.~Mishra, ``An update on ct chest findings in coronavirus
  disease-19 (covid-19),'' \emph{Heart Lung}, vol.~49, no.~5, pp. 442--443,
  2020.

\bibitem{Roberts2021}
M.~Roberts, D.~Driggs, M.~Thorpe, J.~Gilbey, M.~Yeung, S.~Ursprung,
  A.~Aviles-Rivero, C.~Etmann, C.~McCague, L.~Beer, J.~Weir-McCall, Z.~Teng,
  E.~Gkrania-Klotsas, J.~Rudd, E.~Sala, and C.-B. Schönlieb, ``Common pitfalls
  and recommendations for using machine learning to detect and prognosticate
  for covid-19 using chest radiographs and ct scans,'' \emph{Nature Machine
  Intelligence}, vol.~3, pp. 199--217, 2021.

\bibitem{PMSR22}
S.~Reporting, ``Press release: Planqk \& smart reporting - ki-gestützte
  covid-19 befundung von smart reporting evaluiert quantentechnologie für
  schnellere radiologische diagnose,''
  \emph{https://www.smart-reporting.com/de/neuigkeiten/planqk}, 2022.

\bibitem{le2011flexible}
P.~Q. Le, F.~Dong, and K.~Hirota, ``A flexible representation of quantum images
  for polynomial preparation, image compression, and processing operations,''
  \emph{Quantum Information Processing}, vol.~10, pp. 63--84, 2011.

\bibitem{zhang_neqr_2013}
Y.~Zhang, K.~Lu, Y.~Gao, and M.~Wang, ``Neqr: a novel enhanced quantum
  representation of digital images,'' \emph{Quantum Information Processing},
  vol.~12, pp. 2833--2860, 2013.

\bibitem{Jiang_Wang_2014}
\BIBentryALTinterwordspacing
N.~Jiang and L.~Wang, ``\BIBforeignlanguage{en}{Quantum image scaling using
  nearest neighbor interpolation},'' p. 1559–1571, Sep 2014. [Online].
  Available: \url{http://dx.doi.org/10.1007/s11128-014-0841-8}
\BIBentrySTDinterwordspacing

\bibitem{GNEQR}
H.-S. Li, P.~Fan, H.-Y. Xia, H.~Peng, and S.~Song, ``Quantum implementation
  circuits of quantum signal representation and type conversion,'' \emph{IEEE
  Transactions on Circuits and Systems I: Regular Papers}, vol.~66, no.~1, pp.
  341--354, 2019.

\bibitem{QIIP}
B.~{Wang}, M.-q. {Hao}, P.-c. {Li}, and Z.-b. {Liu}, ``{Quantum Representation
  of Indexed Images and its Applications},'' \emph{International Journal of
  Theoretical Physics}, vol.~59, no.~2, pp. 374--402, Nov. 2019.

\bibitem{QBIR}
X.~Liu, D.~Xiao, W.~Huang, and C.~Liu, ``Quantum block image encryption based
  on arnold transform and sine chaotification model,'' \emph{IEEE Access},
  vol.~7, pp. 57\,188--57\,199, 2019.

\bibitem{Wang2019DoubleQC}
L.~Wang, Q.-W. Ran, and J.~Ma, ``Double quantum color images encryption scheme
  based on dqrci,'' \emph{Multimedia Tools and Applications}, vol.~79, pp. 6661
  -- 6687, 2019.

\bibitem{SchuldQMLBook}
M.~Schuld and F.~Petruccione, \emph{Supervised Learning with Quantum
  Computers}.\hskip 1em plus 0.5em minus 0.4em\relax Springer Nature
  Switzerland, 2018.

\bibitem{LaRose2020RobustDE}
R.~LaRose and B.~Coyle, ``Robust data encodings for quantum classifiers,''
  \emph{ArXiv}, vol. abs/2003.01695, 2020.

\bibitem{mitarai2018quantum}
K.~Mitarai, M.~Negoro, M.~Kitagawa, and K.~Fujii, ``Quantum circuit learning,''
  \emph{Physical Review A}, vol.~98, no.~3, p. 032309, 2018.

\bibitem{schuld2020circuit}
M.~Schuld, A.~Bocharov, K.~M. Svore, and N.~Wiebe, ``Circuit-centric quantum
  classifiers,'' \emph{Physical Review A}, vol. 101, no.~3, p. 032308, 2020.

\bibitem{cerezo2021variational}
M.~Cerezo, A.~Arrasmith, R.~Babbush, S.~C. Benjamin, S.~Endo, K.~Fujii, J.~R.
  McClean, K.~Mitarai, X.~Yuan, L.~Cincio \emph{et~al.}, ``Variational quantum
  algorithms,'' \emph{Nature Reviews Physics}, vol.~3, no.~9, pp. 625--644,
  2021.

\bibitem{bergholm2018pennylane}
V.~Bergholm, J.~Izaac, M.~Schuld, C.~Gogolin, S.~Ahmed, V.~Ajith, M.~S. Alam,
  G.~Alonso-Linaje, B.~AkashNarayanan, A.~Asadi \emph{et~al.}, ``Pennylane:
  Automatic differentiation of hybrid quantum-classical computations,''
  \emph{arXiv preprint arXiv:1811.04968}, 2018.

\bibitem{Kingma2014AdamAM}
D.~P. Kingma and J.~Ba, ``Adam: A method for stochastic optimization,''
  \emph{CoRR}, vol. abs/1412.6980, 2014.

\bibitem{Afshar2021}
\BIBentryALTinterwordspacing
P.~Afshar, S.~Heidarian, N.~Enshaei, F.~Naderkhani, M.~J. Rafiee, A.~Oikonomou,
  F.~B. Fard, K.~Samimi, K.~N. Plataniotis, and A.~Mohammadi, ``{COVID-CT-MD,
  COVID-19 computed tomography scan dataset applicable in machine learning and
  deep learning},'' \emph{Scientific Data}, vol.~8, no.~1, p. 121, 2021.
  [Online]. Available: \url{https://doi.org/10.1038/s41597-021-00900-3}
\BIBentrySTDinterwordspacing

\bibitem{GradCAMPlusPlus}
A.~Chattopadhyay, A.~Sarkar, P.~Howlader, and V.~N. Balasubramanian,
  ``Grad-cam++: Generalized gradient-based visual explanations for deep
  convolutional networks,'' \emph{2018 IEEE Winter Conference on Applications
  of Computer Vision (WACV)}, pp. 839--847, 2017.

\bibitem{medmnistv2}
J.~Yang, R.~Shi, D.~Wei, Z.~Liu, L.~Zhao, B.~Ke, H.~Pfister, and B.~Ni,
  ``Medmnist v2-a large-scale lightweight benchmark for 2d and 3d biomedical
  image classification,'' \emph{Scientific Data}, vol.~10, no.~1, p.~41, 2023.

\end{thebibliography}
\begin{figure*}[t] 
\centering
\includegraphics[width=0.74\textwidth]{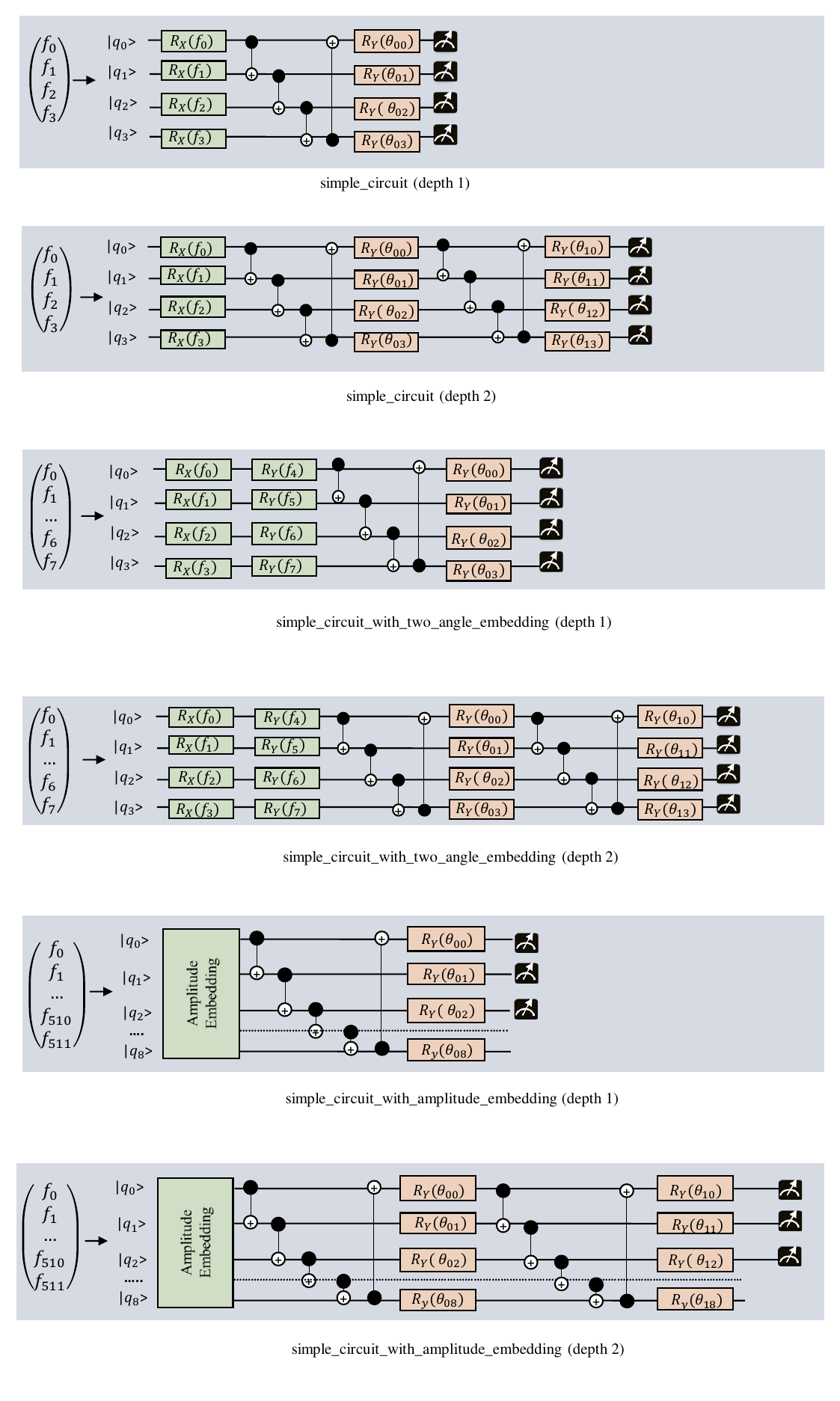}
    \caption {
    Variational quantum circuits utilized in experiments.}
\label{fig:Circuits}
\end{figure*}

\end{document}